\documentclass[final,3p,times]{elsarticle}

\usepackage{color}     
\usepackage{amsmath}   
\usepackage{amssymb}   
\usepackage{amsthm}    
\usepackage{graphicx}  
\usepackage{epsfig}    
\usepackage{epstopdf}  
\usepackage{enumitem}
\usepackage[dvipsnames,table,xcdraw]{xcolor}
\usepackage[ruled, linesnumbered]{algorithm2e}
\usepackage{float}
\usepackage{multirow}
\usepackage{array}
\usepackage{makecell}

\theoremstyle{definition}

\journal{Transportation Research Part C: Emerging Technologies}

\begin{document}

\begin{frontmatter}

\title{
Energy-efficient Reactive and Predictive Connected Cruise Control
}

\author[UoMME]{Minghao Shen\corref{correspauth}}\ead{mhshen@umich.edu}\cortext[correspauth]{Corresponding author.}

\author[GM]{R. Austin Dollar}
\ead{rdollar@clemson.edu}

\author[CaltechME]{Tamas G. Molnar}
\ead{tmolnar@caltech.edu}

\author[PlusAI]{Chaozhe R. He}
\ead{hchaozhe@umich.edu}

\author[ClemsonME]{Ardalan Vahidi}
\ead{avahidi@clemson.edu}

\author[UoMME,UoMCivE]{G\'abor Orosz}
\ead{orosz@umich.edu}

\address[UoMME]{Department of Mechanical Engineering, University of Michigan, Ann Arbor, MI 48109, USA}

\address[GM]{General Motors, Concorde, NC 28027, USA}

\address[CaltechME]{Department of Mechanical and Civil Engineering, California Institute of Technology, Pasadena, CA 91125, USA}

\address[PlusAI]{PlusAI, Inc., Cupertino, CA 95014, USA}

\address[ClemsonME]{Department of Mechanical Engineering, Clemson University, Clemson, SC 29634, USA}

\address[UoMCivE]{Department of Civil and Environmental Engineering, University of Michigan, Ann Arbor, MI 48109, USA}

\begin{abstract}
In this paper, we propose a framework for the longitudinal control of connected and automated vehicles traveling in mixed traffic consisting of connected and non-connected human-driven vehicles. Reactive and predictive controllers are proposed. Reactive controllers are given by explicit feedback control laws. In predictive controllers, the control input is optimized in a receding-horizon fashion, which depends on the predictions of motions of preceding vehicles. Beyond-line-of-sight information is obtained via vehicle-to-vehicle~(V2V) communication, and is utilized in the proposed reactive and predictive controllers. Simulations utilizing real traffic data are used to show that connectivity can bring significant energy savings.
\end{abstract}

\begin{keyword}
connected automated vehicles \sep V2X connectivity \sep MPC \sep traffic flow models
\end{keyword}

\end{frontmatter}


\section{Introduction}\label{sec:intro}

Energy efficiency of vehicles is an everlasting topic in the auto industry, since improving energy efficiency can bring great financial and societal benefits~\cite{ardalan2020book}. As such, driving profiles play an important role in the energy consumption: with the same vehicle traveling on the same route, different drivers may have different driving profiles, which results in great difference in the energy consumption~\cite{saltsman2014impacts}. This shows great potential for improving energy efficiency by optimizing driving profiles.

While human drivers have large variations in their driving behavior~\cite{zarkadoula2007training}, which may undermine the energy efficiency, vehicle automation eliminates such variation and provides a more accurate and consistent way to improve energy efficiency. SAE categorizes automated vehicles into 6 levels (0-5); see Table.~\ref{tab:sae_automation}. Since energy consumption is mainly related to longitudinal motion, level 1-2 automation can already provide significant energy savings. On one hand, automated vehicles~(AV) may optimize their speed profile in advance, taking into consideration the engine and transmission dynamics, and the road elevation along the route~\cite{he2016pcc, he2020improving}. On the other hand, extensive research has focused on optimizing the control input~(pedal, brake and gear shift) to follow some optimal driving cycles~\cite{sciarretta2015ecodrive}. However, these two methods do not take traffic into consideration. In real traffic, vehicles may not be able to follow pre-defined ideal trajectories.

\begin{table}[ht]
\centering
\begin{tabular}{ccccc}
SAE Level &
  \thead{Execution of steering,\\ acceleration/deceleration} &
  \thead{Monitoring of\\driving environment} &
  \thead{Fallback performance\\ of dynamic driving task} &
  \thead{System capability\\(driving modes)} \\ \hline
0~(No automation)         & Human            & Human & Human & N/A                \\
1~(Driver assistance)      & Human \& system & Human & Human & Some driving modes \\
2~(Partial automation)     & System                  & Human & Human & Some driving modes \\
3~(Conditional automation) & System                  & System       & Human & Some driving modes \\
4~(High automation)        & System                  & System       & System       & Some driving modes \\
5~(Full automation)      & System                  & System       & System       & All driving modes 
\end{tabular}%
\caption{SAE Levels of Vehicle Automation}
\label{tab:sae_automation}
\end{table}

\begin{table}[h!]
\centering
\begin{tabular}{cccc}
                  & \thead{Motion Prediction \\ of Remote Vehicle}    & \thead{Motion Planning \\of Ego Vehicle}      & \thead{Control of \\ Ego Vehicle} \\ \hline
Status Sharing           & Ego Vehicle    & Ego Vehicle    & Ego Vehicle    \\
Intent Sharing    & Remote Vehicle & Ego Vehicle                            & Ego Vehicle               \\
Agreement Seeking & Remote Vehicle & Remote Vehicle & Ego Vehicle               \\
Prescriptive Cooperation & Remote Vehicle & Remote Vehicle & Remote Vehicle
\end{tabular}
\caption{SAE Levels of Cooperative Driving Automation~(CDA)}
\label{tab:saecda}
\end{table}

With level 1 or 2 automation, AVs rely on \emph{adaptive cruise control~(ACC)} algorithm to react to motion of preceding vehicles in traffic. ACC has a long history tracing back to the 1990s. The controller design usually falls into one of the two categories: \emph{reactive controller} or \emph{predictive controller}. Reactive ACC~(RACC) has explicit feedback control laws that are usually parameterized, so that the controller parameters can be optimized for the energy efficiency while ensuring other specifications such as stability. On the other hand, predictive ACC~(PACC) can directly optimize the future trajectory based on the prediction of the future motions of neighboring vehicles. While predictions may significantly improve energy efficiency, an accurate prediction is very hard to make without additional information, since the motions of neighboring vehicles can be highly correlated or completely stochastic. 

Vehicle-to-vehicle~(V2V) communication can potentially resolve this problem. Peer-to-peer communication enables connected vehicles to share information for prediction and control, as well as facilitates cooperation among vehicles in the traffic. SAE categorized cooperative driving automation~(CDA) into status-sharing, intent-sharing, agreement-seeking and prescriptive cooperation~\cite{sae2021CDA}; see Table~\ref{tab:saecda}. Many of the existing research works assume high level of cooperation, e.g., prescriptive cooperation. Assuming that an entire platoon of vehicles are connected and automated, the high level of cooperation enables centralized control over all these vehicles. Such controllers are often referred to as \emph{cooperative adaptive cruise control~(CACC)}. Similar to ACC, CACC design can also be categorized into reactive and predictive control~\cite{berkeley2015CACC,barth2018reviewCACC}.  Reactive control tries to synchronize the speed of the platoon, guaranteeing string stability and maintaining desirable headway~\cite{VanDeWouW2014stringstb, yangzheng2018hinfPlatoon}. On the other hand, predictive controllers have access to the future motion plans of leading vehicles, therefore coordinated and even global optimization becomes possible~\cite{shengbo2015TopoPlatoon,shengbo2017MPCplatoon,wouw2019MPC, borrelli2018CAVsurvey, Johansson_predictive_framework}. To make the system more scalable, distributed control protocol has also been studied~\cite{crowley2020DMPCplatoon}. Research has shown that CACC and platooning bring significant energy benefits under different scenarios~\cite{borrelli2017ecoCACC,shengbo2021distrEcoMPC,shengbo2018Urban}. However, currently the V2V technology is far from being widely deployed. The assumption of high penetration of connectivity and high level of cooperation is hard to realize in practice in the near future.

\begin{figure}[h!]
    \centering
    \includegraphics[width=0.5\textwidth]{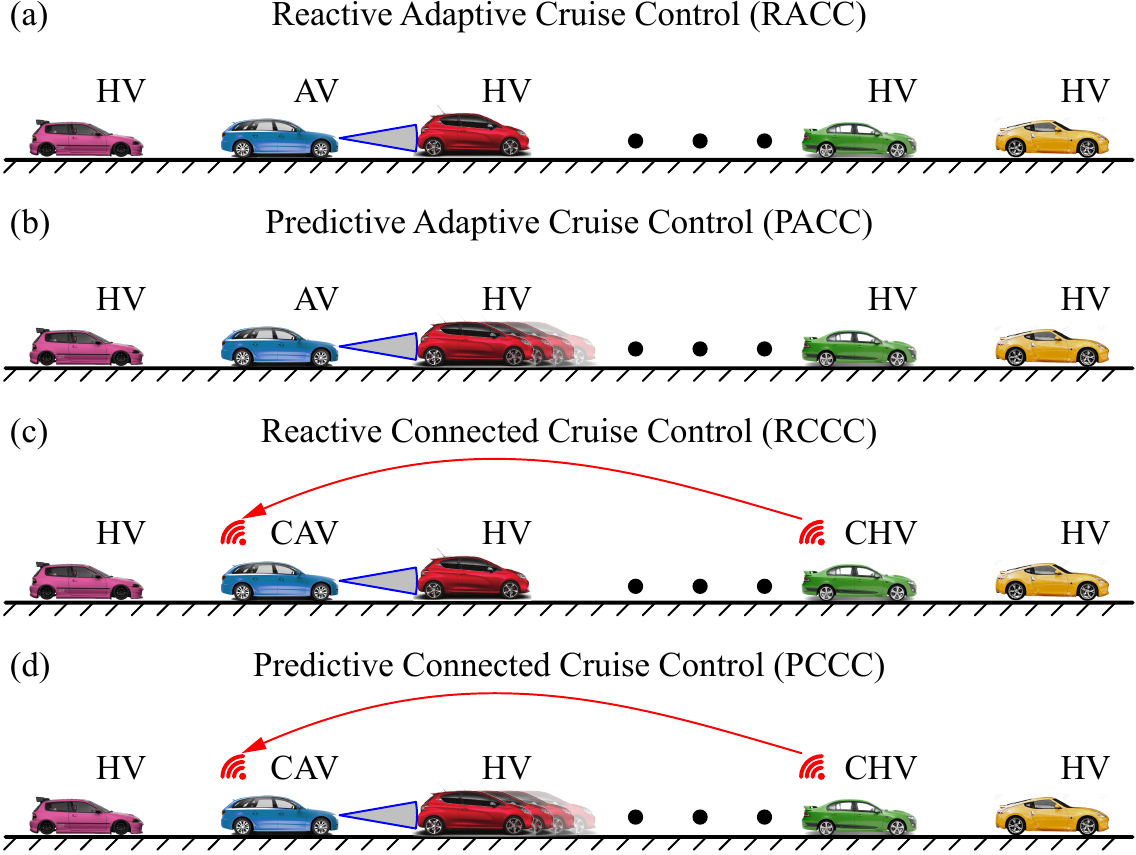}
    \caption{Illustration of longitudinal control strategies for automated vehicles~(AVs) and connected automated vehicles~(CAVs) traveling in mixed traffic that involves human-driven vehicles~(HVs) and connected human-driven vehicles~(CHVs). Predictive controllers rely on the predictions on the future motions of preceding vehicles, as is shown in shadowed vehicles.}
    \label{fig:Concepts}
\end{figure}

\begin{figure}[ht]
    \centering
    \includegraphics[width=0.6\textwidth]{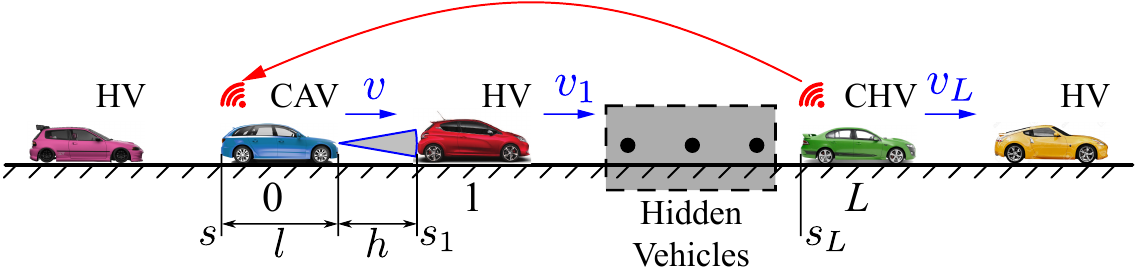}
    \caption{Connected cruise control in mixed traffic consisting of connected and non-connected vehicles.}
    \label{fig:ccc_nofun}
\end{figure}

The near future of transportation is more likely to evolve into mixed traffic. Controllers that operate under mixed traffic consisting of connected and non-connected vehicles are referred to as \emph{connected cruise control~(CCC)}. Only low-level cooperation as status-sharing is assumed and centralized control is not possible. Potentially four kinds of vehicles may paticipate in the mixed traffic: human-driven vehicle~(HV), connected human-driven vehicle~(CHV), automated vehicle~(AV) and connected and automated vehicle~(CAV). Without connectivity, the longitudinal controllers for AVs execute adaptive cruise control. While with connectivity, CAVs may execute more performant controllers, even with low level of cooperation such as status-sharing protocol. In CCC, CAVs have access to beyond-line-of-sight information of CHVs and CAVs in the distance, which is incorporated into the controller design. Similar to ACC and CACC, CCC can also be categorized into reactive control and predictive control. Reactive CCC~(RCCC) takes the V2V information from leading vehicles as reference signals, the objective is still to synchronize the speed in the traffic for string stability and smooth driving~\cite{gabor2016ccc,Linjun2016Motif, Linjun2017consensus}. Meanwhile, predictive CCC~(PCCC) can incorporate the information of preceding vehicles to make predictions on the motion of the vehicle immediately in the front~\cite{austin2021PCCC}. This may significantly improve predictions, and enable optimized planning of motions in advance, which may reduce speed variations and save energy. In Fig.~\ref{fig:Concepts} the concepts of RACC, PACC, RCCC, and PCCC are illustrated graphically for mixed traffic scenarios containing HVs, AVs, CHVs, and CAVs. 

With all these distinctions made, this paper presents contributions to improve energy efficiency in mixed traffic as follows:
\begin{itemize}
\item We provide design framework on reactive and predictive control of connected automated vehicles driving in mixed traffic consisting of connected and non-connected vehicles.
\item Under both reactive and predictive controller framework, we show the significant energy benefits provided by V2V connectivity. We provide explanations to the energy savings by comparing simulated trajectories.
\item We compare the reactive and predictive controllers in three typical scenarios and show the benefit of predictive controllers while utilizing real traffic data. 
\end{itemize}

The remainder of this paper is organized as follows. Section~\ref{sec:vehdyn} introduces the problem setting and longitudinal dynamics of vehicles. Section~\ref{sec:reactive_controller} discusses the design of energy-efficient reactive controllers including RACC and RCCC. Section~\ref{sec:predictive_controller} discusses predictive controller designs including PACC and PCCC. Section~\ref{sec:num_result} shows the energy benefits of different controller designs with lean penetration of connected vehicles. Section~\ref{sec:concl} concludes this paper and points out future research directions.

\section{Vehicle Dynamics}\label{sec:vehdyn}

In this section, we introduce the problem setup, and derive the state space model for longitudinal controller design. Consider the connected cruise control scenario in Fig.~\ref{fig:ccc_nofun}, in which a connected and automated vehicle~(CAV) is driving on a flat road without elevation change, with the intention to follow human-driven traffic. The longitudinal dynamics of the CAV with respect to its position $s$ and velocity $v$ can be modeled as in~\cite{orosz2020survey}:
\begin{equation}\label{eqn:veh_dyn_original}
\begin{aligned}
\dot{s} &= v~, \\
\dot{v} &= -\frac{1}{m_{\mathrm{eff}}}\Big(mg\xi + kv^2\Big) + \frac{T_{\mathrm{w}}}{m_{\mathrm{eff}}R}~.
\end{aligned}
\end{equation}
Here the effective mass ${m_{\mathrm{eff}} = m + I/R^2}$ incorporates the mass $m$, mass moment of inertia $I$ and the radius $R$ of the wheels. Moreover, $g$ is the gravitational constant,  $\xi$ denotes the rolling resistance coefficient and $k$ denotes the air resistance coefficient. We can control the vehicle speed by applying different torque on wheels $T_{\mathrm{w}}$ using the engine/electric motors and the brakes. To highlight how control actions influence the system, we consider the commanded acceleration as control input $u$, and rewrite \eqref{eqn:veh_dyn_original} as
\begin{equation}\label{eqn:simple_nonlin_dyn}
\begin{aligned}
\dot{s}(t) &= v(t)~, \\
\dot{v}(t) &= -f\Big(v(t)\Big) + \mathrm{sat}\Big(u(t-\sigma)\Big)~,
\end{aligned}
\end{equation}
where
\begin{equation}\label{eqn:f_satu_def}
f(v) = -\frac{1}{m_{\mathrm{eff}}}\left(mg\xi + kv^2\right)~, \quad \mathrm{sat}\Big(u(t-\sigma)\Big) = \frac{T_{\mathrm{w}}(t)}{m_{\mathrm{eff}}R}~.
\end{equation}
The model incorporates the delay $\sigma$ in powertrain system, and the saturation $\mathrm{sat}(\cdot)$ arising from limitations of engine/motor power, engine/motor torque and braking capability. More specifically, the saturation is modeled as
\begin{equation}\label{eqn:sat_def}
\mathrm{sat}(u) = \min\Big\{\tilde{u}_{\max}, \max\{u_{\min}, u\}\Big\}~,
\end{equation}
\begin{equation}\label{eqn:umax_def}
\tilde{u}_{\max} = \min\left\{u_{\max},~m_1 v + b_1, ~ m_2 v + b_2\right\}~,
\end{equation}
as is shown in Fig.~\ref{fig:ovm_fun}(a) and (b). Here $u_{\min}$ is the minimum acceleration~(maximum deceleration) due to the braking capability, and $m_1$, $m_2$, $b_1$, $b_2$ are determined by engine torque limit and power limit.

In order to follow the desired acceleration, $a_{\mathrm{d}}$, the control action
\begin{equation}\label{eqn:ucomp_ftil}
u(t) = \tilde{f}\Big(v(t)\Big) + a_{\mathrm{d}}(t)~,
\end{equation}
is applied, where the term $\tilde{f}$ tries to compensate the nonlinear physical effects $f$ in \eqref{eqn:f_satu_def}. In this article, we assume that perfect compensation is possible and focus on the choice of desired acceleration $a_{\mathrm{d}}$, which simplifies~\eqref{eqn:veh_dyn_original} to
\begin{equation}\label{eqn:vehdyn_ideal}
\begin{aligned}
\dot{s}(t) &= v(t)~, 
\\
\dot{v}(t) &= \mathrm{sat}\Big(a_{\mathrm{d}}(t-\sigma)\Big)~,
\end{aligned}
\end{equation}

Energy consumption is the main interest in this article. It is evaluated with energy consumption per unit mass
\begin{equation}
w = \int_{t_0}^{t_f} v(t)g\Big(\dot{v}(t) + f(v(t))\Big)\mathrm{d}t~,
\end{equation}
where $g(x)=\max\{x, 0\}$ implies that braking does not consume or recover energy. We remark that the effects of energy recovering systems can be included by choosing different $g$ functions, but this is beyond the scope of this article.

In what follows, we investigate the energy efficiency of four types of controllers: RACC, RCCC, PACC and PCCC; as summarized by Table~\ref{tab:alg_comp}. These four controllers are detailed in the next two sections and in Algorithms~1-4.

\section{Reactive Controllers}\label{sec:reactive_controller}

\begin{figure}[h!]
    \centering
    \includegraphics[width=\textwidth]{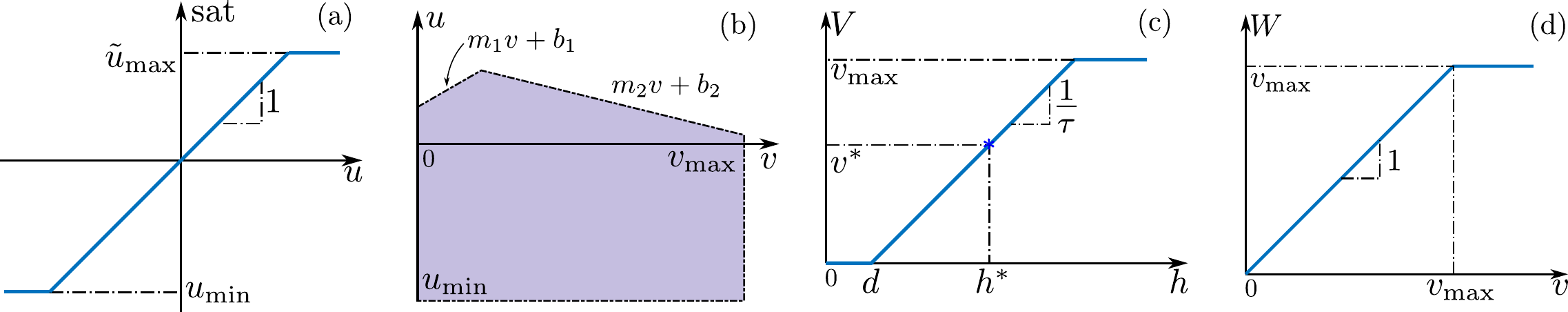}
    \caption{Nonlinearities in vehicle dynamics and optimal velocity model~(OVM). (a) Saturation function~\eqref{eqn:sat_def}. (b) Acceleration limits~\eqref{eqn:umax_def}. (c) Range policy~\eqref{eqn:range_policy}. (d) Speed policy~\eqref{eqn:speed_policy}.}
    \label{fig:ovm_fun}
\end{figure}

In this section, we design control algorithms for reactive adaptive cruise control~(RACC) and reactive connected cruise control~(RCCC). We start with the simple RACC case, where an automated vehicle is controlled and there is no connected vehicle in the traffic, as is shown in Fig.~\ref{fig:Concepts}(a). With on-board sensors such as camera, lidar or radar, the ego vehicle can only react to the vehicle immediately in the front. RACC determines the desired acceleration $a_{\mathrm{d}}$ as a function of headway~$h$, its speed~$v$, as well as the speed~$v_1$ of the vehicle immediately in the front:
\begin{equation}
a_{\mathrm{d}} = F(h, v, v_1)~,
\end{equation}
where ${h = s_1 - s - l}$ is related to the positions $s$ and $s_1$ of the vehicles and the length~$l$ of the ego vehicle, as is shown in Fig.~\ref{fig:ccc_nofun}. For example, optimal velocity model~(OVM) yields the control algorithm
\begin{equation}\label{eqn:OVM}
F^{\mathrm{OVM}}(h, v, v_1) = \alpha \Big(V(h) - v\Big) + \beta \Big(W(v_1) - v\Big)~,
\end{equation}
where the range policy $V(h)$ determines the desired velocity for headway and the speed policy $W(v_1)$. A common choice of range policy is given in~\cite{gabor2016ccc} as
\begin{equation}\label{eqn:range_policy}
V(h) = \min\Big\{v_{\max}, \max\big\{0, (h-d) / \tau\big\}\Big\}~.
\end{equation}
As is shown in Fig.~\ref{fig:ovm_fun}(c), when the headway is less than the stopping distance $d$, the ego vehicle tends to stay still, while when the headway is larger than $d+ \tau v_{\max}$, the ego vehicle intends to travel with maximum speed $v_{\max}$ while not being influenced by the preceding vehicle. For intermediate headway distance, the desired velocity grows with constant gradient $1/\tau$ where $\tau$ is referred to as \emph{time headway}. Moreover, the speed policy 
\begin{equation}\label{eqn:speed_policy}
W(v_1) = \min\big\{v_{\max},v_1\big\}~.
\end{equation}
is used to prevent the ego vehicle from speeding once the preceding vehicle goes faster than $v_{\max}$; see Fig.~\ref{fig:ovm_fun}(d).

Another widely-used car-following model is the intelligent driver model~(IDM)~\cite{treiber2000IDMpaper}:
\begin{equation}\label{eqn:IDM_law}
F^{\mathrm{IDM}}(h, v, v_1) =  a_0\left(1-\left(\frac{v}{v_{\max}}\right)^\delta - \left(\frac{H(v, v_1)}{h}\right)^2\right)
\end{equation}
where the desired headway is calculated using the range policy
\begin{equation}\label{eqn:IDM_hidm}
H(v, v_1) = d + \max\left\{0, \tau v - \frac{v(v_1-v)}{\sqrt{a_0b_0}}\right\}~.
\end{equation}
Here $a_0$ corresponds to the maximum acceleration, $b_0$ is the deceleration coefficient, $\tau$ is the desired time headway. Note that at steady state~($v_1 = v$), the last term is eliminated and we obtain the simplified range policy
\begin{equation}\label{eqn:idm_range_simplified}
H(v) = d + \tau v~,
\end{equation}
which is the inverse of $V(h)$ in \eqref{eqn:range_policy} when ${0 < v < v_{\max}}$.
Stopping distance $d$ and maximum speed $v_{\max}$ have the same meaning as those in \eqref{eqn:range_policy}.

When V2V connectivity is available, connected and automated vehicles may rely on information from connected human driven vehicles and execute~\emph{reactive connected cruise control~(RCCC)}; see Fig.~\ref{fig:Concepts}(c). In this case the ego vehicle not only reacts to the vehicle immediately in the front, but also to the vehicles beyond line of sight.
\begin{equation}\label{eqn:ad_rccc_general}
a_{\mathrm{d}} = F\Big(h, v, \{s_i\}_{i\in\mathcal{I}}~. \{v_{i}\}_{i\in \mathcal{I}}\Big)~,
\end{equation}
Here $\mathcal{I}$ denotes the set of all the vehicles that are connected to or sensed by the ego vehicle, so $1\in\mathcal{I}$ because even if vehicle $1$ is not connected vehicle, it can still be sensed by ego vehicle using onboard sensors. For example, one can extend OVM \eqref{eqn:OVM} to the RCCC controller
\begin{equation}\label{eqn:rccc_nodelay}
F\Big(h, v, \{v_{i}\}_{i\in \mathcal{I}}\Big) = \alpha\Big(V(h)-v\Big) + \sum_{i\in \mathcal{I}}\beta_i\Big(W(v_i) - v\Big)~.
\end{equation}
Notice that, the signals $\{v_i\}_{i\in \mathcal{I}}$ are reference signals. The objective of our control design is to minimize speed variation while maintaining a reasonable headway. The controller does not necessarily need to respond to every reference signal immediately. Instead, the controller may wait for a while before responding to the signals from vehicles in the distance~\cite{minghao2021delayCCC}. Thus, \eqref{eqn:rccc_nodelay} can be generalized to
\begin{equation}\label{eqn:rccc_delay}
F(t) = \alpha\Big(V(h(t)) - v(t)\Big) + \sum_{i\in \mathcal{I}}\beta_i \Big(W(v_i(t-\sigma_i)) - v(t)\Big)~,
\end{equation}
where the delay $\sigma_i$-s are additional design parameters, as oppose to $\sigma$ in~\eqref{eqn:vehdyn_ideal} which is a result of powertrain dynamics. In this paper, we refer to this as \emph{reactive connected cruise control(RCCC)}. Notice that, RACC is essentially a special case of RCCC, where $\beta_i = 0$ for all $i \in \mathcal{I}\setminus\{1\}$. The RACC and RCCC algorithms are summarized in Algorithm~\ref{alg:RACC} and Algorithm~\ref{alg:RCCC}, respectively. There has been extensive research on the choice of energy-optimal controller parameters of RACC and RCCC; we refer to~\cite{minghao2022stochasticCCC} for more details.

We remark that safety is often specified as maintaining larger headway than the velocity-dependent miminum headway, e.g., ${H_{\rm min}(v) = d_{\rm min} + \tau_{\rm min} v}$, cf.~\eqref{eqn:idm_range_simplified}. In order to ensure that an RACC or a RCCC controller is safe, one may select the controller parameters such that a so-called barrier function maintains positive derivative~\cite{chaozhe2018}. Alternatively, one may use so-called control barrier functions to modify RACC and RCCC controllers and make them safe~\cite{ames2017ACCCBF,anil2022CBF}. Such controller tunings and modifications are beyond the scope of this paper and are not discussed here in details.

\begin{algorithm}[h!]
\SetKwInput{kwInit}{Initialize}
\For{$t=0$ \emph{\KwTo} $T_{\max}$}
{Observe and store $s_1(t)$, $v_1(t)$\;

Calculate desired acceleration $a_{\mathrm{d}}$ using~\eqref{eqn:OVM}\;
Apply control command $u = \tilde{f}(v) + a_{\mathrm{d}}$\;
}
\caption{Reactive Adaptive Cruise Control~(RACC)}
\label{alg:RACC}
\end{algorithm}

\begin{algorithm}[h!]
\SetKwInput{kwInit}{Initialize}
\For{$t=0$ \emph{\KwTo} $T_{\max}$}
{Observe and store $s_1(t)$, $v_i(t)$, $i\in \mathcal{I}$\;
Calculate desired acceleration $a_{\mathrm{d}}$ using~\eqref{eqn:rccc_delay}\;
Apply control command $u = \tilde{f}(v) + a_{\mathrm{d}}$\;
}
\caption{Reactive Connected Cruise Control~(RCCC)}
\label{alg:RCCC}
\end{algorithm}

\section{Predictive Controllers}\label{sec:predictive_controller}

Apart from constructing explicit reactive control laws, one may also formulate control synthesis as an optimization problem, in which the objective functions and constraints are utilized so that control actions optimize the performance. Model predictive control~(MPC) is a prevailing choice, which relies on predicting the motion of the vehicle in front of the ego vehicle. An accurate prediction is crucial, for both predictive adaptive cruise control~(PACC) illustrated in Fig.~\ref{fig:Concepts}(b) and the predictive connected cruise control~(PCCC) illustrated in Fig.~\ref{fig:Concepts}(d). One needs to make prediction on the future trajectory of preceding vehicle, and choose optimal action based on the prediction. The choice of predictor significantly influences the energy efficiency and safety of predictive controllers. With V2V communication, the CAV gets access to information from vehicles ahead in the distance, which enables better prediction of the vehicle immediately in the front. Despite the change of connectivity structure, the general optimization formulation remains the same for both PACC and PCCC. Only the estimation on the position of preceding vehicle $\hat{s}_1$ differs in different connectivity structures.

Let ${x=[s,\ v]^\top}$ represent the state of the ego vehicle and ${x_1=[s_1, \ v_1]^\top}$ represent the state of the vehicle immediately in the front. At time $t$, the system specification can be formulated into the continuous-time optimization problem
\begin{equation}\label{eqn:continuous_opt}
\begin{aligned}
\min ~ & \int_{t}^{t+T} \ell \Big( x\big(\tilde{t}|t\big), \hat{x}_1\big(\tilde{t}|t\big), a\big(\tilde{t}|t\big) \Big) \mathrm{d}\tilde{t}~,
\\
\mathrm{s.t.}~ & G_{\mathrm{dynamics}}\Big( x\big(\tilde{t}|t\big), a\big(\tilde{t}|t\big) \Big) = 0~,
\\
& G_{\mathrm{safety}}\Big( x\big(\tilde{t}|t\big), \hat{x}_1\big(\tilde{t}|t\big) \Big) \leq 0~, 
\\
& G_{\mathrm{saturation}}\Big( x\big(\tilde{t}|t\big), a\big(\tilde{t}|t\big) \Big) \leq 0~, 
\\
& x\big(t|t\big) = x(t)~, 
\\
&a\big(\tilde{t}|t\big) = a_{\mathrm{d}}\big(\tilde{t}-\sigma\big)~, \quad \forall \tilde{t}\in [t, t+\sigma)~.
\end{aligned}
\end{equation}
That is, under safety constraints $G_{\mathrm{safety}}$ and saturation $G_{\mathrm{saturation}}$, we aim to minimize the cumulative cost function $\ell$ during the time interval ${[t, t+T]}$ based on our knowledge of the system behavior $x\big(\tilde{t}|t\big)$, and our prediction on the future motion of preceding vehicle $\hat{x}_1\big(\tilde{t}|t\big)$. Due to the powertrain delay, when ${\tilde{t}\leq t+\sigma}$, the acceleration at time $\tilde{t}$ is determined by the desired acceleration at ${\tilde{t}-\sigma}$. So the control input at time $t$ shall determine the acceleration at time ${t+\sigma}$. In other words, ${a_{\mathrm{d}}(t) = a(t+\sigma|t)}$.

The objective function of the MPC controller includes penalties on the headway and the control input. Similar to the range policies~\eqref{eqn:range_policy}, \eqref{eqn:IDM_hidm}, \eqref{eqn:idm_range_simplified} used in the reactive controllers, the predictive controller aims to keep a desirable headway as a function of velocity. For example, in this paper, we aim to keep a constant time headway $\tau$ and thus we utilize \eqref{eqn:idm_range_simplified}.
As shown below, the MPC controller typically applies a quadratic penalty on the deviation of headway from desirable values. In addition, we will also penalize the magnitude of the control input as given below.

While the optimization problem~\eqref{eqn:continuous_opt} is defined in continuous time, for efficient implementation, we usually need to convert it to discrete time optimization. We first transform the dynamics \eqref{eqn:vehdyn_ideal} into discrete time using the time step $\Delta t$. To make the final MPC a convex quadratic programming~(QP) problem, we drop the nonlinear terms and move the saturation function to inequality constraints. Thus, the equality constraints are given by the linear dynamics
\begin{equation}\label{eqn:disc_cardyn}
\begin{aligned}
s(k+1) &= s(k) + \Delta t~v(k) + \frac{1}{2}\Delta t^2 a(k)~, 
\\
v(k+1) &= v(k) + \Delta t~a(k)~,
\end{aligned}
\end{equation}
while the saturation function \eqref{eqn:sat_def} is transformed to the inequality constraints
\begin{equation}
a(k) \geq u_{\min}~, \quad 
a(k) \leq m_1 v(k) + b_1~, 
\quad a(k) \leq m_2 v(k) + b_2~.
\end{equation}
Due to powertrain delay $\sigma$, the acceleration at current time is determined by the control input in the past, that is,
\begin{equation}\label{eqn:disc_delay}
a(k) = a_{\mathrm{d}}(k-q)~,
\end{equation}
where ${\sigma=q\Delta t}$. 
In addition, we define a minimum headway to guarantee safety
\begin{equation}
H_{\min}(v) = d_{\min} + \tau_{\min}v~.
\end{equation}
In order to compensate for the prediction uncertainty, we impose additional safety-margin $d_{\mathrm{margin}}(k)$ at each time $k$ in prediction horizon,
\begin{equation}\label{eqn:safety_ineq}
\hat{h}(k) - H_{\min}\big(v(k)\big) - d_{\mathrm{margin}}(k) \geq 0~,
\end{equation}
where ${\hat{h}(k) = \hat{s}_1(k) - s(k) - l}$ is the estimated headway. The derivation of the safety margin is elaborated in \ref{sec:margin}.

In summary, the MPC controller is formulated as follows
\begin{equation}\label{eqn:mpc_discrete}
\begin{aligned}
\min_{a(0|t), \ldots, a(T-1+q|t), \epsilon}~& q_{\mathrm{g}} \sum_{k=0}^{T} \left(\hat{h}(k|t) - H\big(v(k|t)\big)\right)^2 
+  q_{\mathrm{a}} \sum_{k=0}^{T-1} a^2(k|t) + q_{\epsilon} \epsilon~, 
\\
\mathrm{s.t.} \quad & s(k+1|t) = s(k|t) + \Delta t~v(k|t) + \frac{1}{2}\Delta t^2~a(k|t)~, \quad k =0,\ldots, T-1~,
\\
& v(k+1|t) = v(k|t) + \Delta t~a(k|t)~, \,\,\, \qquad \qquad \qquad k = 0, \ldots, T-1~,
\\
& \hat{h}(k|t) = \hat{s}_1(k|t) - s(k|t) - l~, \,\,\, \quad \qquad \qquad \qquad k=0,\ldots, T~, 
\\
& \hat{h}(k|t) - H_{\min}\big(v(k|t)\big) - d_{\mathrm{margin}}(k) \geq -\epsilon~, \quad \quad \quad k = 0, \ldots, T~,
\\
& 0\leq v(k|t) \leq v_{\max}~, \,\quad \qquad \qquad \qquad \qquad \qquad  k = 0, \ldots, T~, 
\\
&u_{\min}\leq a(k|t)~, \,\,\,\,\, \qquad \qquad \qquad \qquad \qquad \qquad k=0,\ldots, T~,
\\
&a(k+q|t) \leq m_1 v(k+q|t) + b_1~, \, \qquad \qquad \qquad k=0,\ldots, T~,
\\
& a(k+q|t) \leq m_2 v(k+q|t) + b_2~, \, \qquad \qquad \qquad k=0,\ldots, T~,
\\
& s_1(0|t) = s_1(t)~, \quad s(0|t) = s(t)~, \quad v(0|t) = v(t)~, 
\\
& a(k|t) = a_{\mathrm{d}}(t + k -q)~, \,\,\, \quad \qquad \qquad \qquad \qquad  k = 0, \ldots, q-1~,
\end{aligned}
\end{equation}
where, by abuse of notation, $t$ and $T$ represent discrete time and we apply soft constraint with ${\epsilon \geq 0}$ in the safety inequality~\eqref{eqn:safety_ineq} to ensure feasibility.

\subsection{Predictive Adaptive Cruise Control~(PACC)}

Accurate prediction of the preceding vehicles is the key to the success of MPC controller. In PACC, no extra information on the preceding vehicle is available from V2V connectivity, it is a common choice to assume that the preceding vehicle maintains its the current speed in the future~\cite{chaozhe2017MPC,chaozhe2020fuel}:
\begin{equation}
\begin{aligned}
\hat{s}_1\big(\tilde{t}|t\big) &= s_1(t) + \big(\tilde{t} - t\big)~v(t)~, \quad  \tilde{t}\geq t~, 
\\
\hat{v}_1\big(\tilde{t}|t\big) &= v_1(t)~, \qquad \qquad \qquad  \tilde{t} \geq t~.
\end{aligned}
\end{equation}
The corresponding PACC algorithm, which is in discrete time, is shown in Algorithm~\ref{alg:PACC}.
With connectivity, PCCC may leverage additional information about the preceding vehicle that may potentially bring more energy benefits.

\begin{algorithm}
\For{$t=0$ \emph{\KwTo} $T_{\max}$}
{Observe and store $s_1(t)$, $v_1(t)$\;
Apply constant speed prediction on leading vehicle $1$
$${\hat{s}_1(t+k|t) = s_1(t) + k\Delta t~v_1(t)~, \quad  k = 1,\ldots, T~,\;}$$
$${\hat{v}_1(t+k|t) = v_1(t)~, \,\,\, \quad \qquad \qquad k = 1,\ldots, T~,\;}$$

Solve the optimization problem~\eqref{eqn:mpc_discrete} using $\hat{s}_1(t:t+T|t)$\;
Output desired acceleration $a(q)$ to powertrain and braking systems\;
Update with discrete time dynamics~\eqref{eqn:disc_cardyn}, \eqref{eqn:disc_delay}\;
}
\caption{Predictive Adaptive Cruise Control~(PACC)}
\label{alg:PACC}
\end{algorithm}
\vspace{-3mm}

\subsection{Predictive Connected Cruise Control~(PCCC)}
\label{sec:pccc}

In this section, we introduce predictive connected cruise control which utilizes V2V connectivity when available. We consider the scenario in Fig.~\ref{fig:Concepts}(d) where a CAV executes PCCC. We assume lean penetration of connectivity where only a single lead CHV~(vehicle~$L$) is connected to the CAV while the preceding vehicle~(vehicle~$1$) is sensed by on-board sensors.

\begin{algorithm}
\For{$t=0$ \emph{\KwTo} $T_{\max}$}
{Observe and store $s_1(t)$, $v_1(t)$, $s_L(t)$, $v_L(t)$\;
Estimate $\hat{n}_{\mathrm{h}}$\;
Apply constant speed prediction on leading vehicle $L$
$${\hat{s}_L(t+k|t) = s_L(t) + k\Delta t~ v_L(t)~, \quad k = 1,\ldots, T~,\;}$$
$${\hat{v}_L(t+k|t) = v_L(t)~, \,\,\, \quad \qquad \qquad k = 1,\ldots, T~,\;}$$

\eIf{$\hat{n}_{\mathrm{h}}==0$}
{Simulate ${\hat{s}_1(t:t+T|t)}$ following the vehicle numbered $L$\;}
{\For{$i=0$ \emph{\KwTo} $\hat{n}_{\mathrm{h}}$}
{Simulate ${\hat{s}_{L-i-1}(t:t+T|t)}$ following vehicle ${L-i}$ using uniform flow initialization\;}
Simulate ${\hat{s}_{1}(t:t+T|t)}$ following vehicle $2$ using real data initialization\;}
Solve the optimization problem~\eqref{eqn:mpc_discrete}\;
Output desired acceleration $a(q)$ to powertrain and braking systems\;
Update with discrete time dynamics~\eqref{eqn:disc_cardyn}, \eqref{eqn:disc_delay}\;
}
\caption{Predictive Connected Cruise Control~(PCCC)}
\label{alg:PCCC}
\end{algorithm}

We propose the PCCC control framework detailed Algorithm~\ref{alg:PCCC}. Compared to the PACC in Algorithm~\ref{alg:PACC}, the only change is in the way we predict ${\hat{s}_1(t: t+T)}$. In PACC, without additional information, we chose to make constant speed assumption. While in PCCC, with the additional information from V2V communication, we can potentially make more accurate prediction on the future motion of preceding vehicles. In this paper, we apply constant speed assumption on the future motion of the connected vehicle in the distance, then we simulate the motion of subsequent vehicles, until reaching the vehicle immediately ahead. In traffic with lean penetration of connected vehicles, the number $\hat{n}_{\mathrm{h}}$ of hidden vehicles driving between the vehicle immediately ahead~(vehicle~$1$) and the connected vehicle in the distance~(vehicle~$L$) is unknown; see Fig.~\ref{fig:ccc_nofun}. In the next subsection, we introduce an algorithm to estimate the number of hidden vehicles. We summarize four controllers in Table~\ref{tab:alg_comp}.

\begin{table}[h!]
\centering
\begin{tabular}{cccc}
     & Explicit Law & Connectivity & Prediction \\ \hline
RACC & Yes          & No           & No              \\
RCCC & Yes          & Yes          & No              \\
PACC & No           & No           & Constant Speed  \\
PCCC & No           & Yes          & Car-following Model 
\end{tabular}
\caption{Comparison of cruise control algorithms}
\label{tab:alg_comp}
\end{table}

\subsection{Hidden Vehicle Estimation}
\label{subsec:hv_est}
The estimation of the number of hidden vehicles is based on historical data of preceding vehicles: $s_1$, $v_1$, $s_L$ and $v_L$, which are recorded during driving. We denote the resulting estimation by $\hat{n}_{\mathrm{h}}$ that approximates the unknown number of hidden vehicles, ${n_{\mathrm{h}} = L - 2}$.

The detailed algorithm design is shown in Algorithm~\ref{alg:nh_est}. Our estimation algorithm conducts brute-force search for every possible number of hidden vehicle $n_{\mathrm{h}}$. At time $t$ for a given $n_{\mathrm{h}}$, we consecutively simulate the motion of hidden vehicles over the past $[t-T_{\mathrm{h}}, t]$ using the IDM model~\eqref{eqn:IDM_law}, \eqref{eqn:IDM_hidm}, to obtain an estimation $\hat{s}_1^{(n_{\mathrm{h}})}$ of the position $s_1$ of the preceding vehicle. Then we compare the simulated $s_1^{(n_{\mathrm{h}})}$ with the recorded measurements of $s_1$. The $n_{\mathrm{h}}$ corresponding to the minimal error is chosen.

\begin{algorithm}[h!]
\SetKwInOut{KwInput}{Input}
\SetKwInOut{KwOutput}{Output}
\KwInput{History trajectories $s_{1}(t-T_{\mathrm{h}}:t)$, $s_{L}(t-T_{\mathrm{h}}:t)$, $v_{1}(t-T_{\mathrm{h}}:t)$, $v_{L}(t-T_{\mathrm{h}}:t)$; Current speed $v(t)$\;Previous estimation $\hat{n}_{\mathrm{h}}^{\mathrm{prev}}$\;}
\KwOutput{Estimated number of hidden vehicles $\hat{n}_{\mathrm{h}}$\;}
$n_{\mathrm{h, max}} = \min\left\{\hat{n}_{\mathrm{h}}^{\mathrm{prev}} + 1, ~\left \lceil\frac{s_L(t) - s_1(t)}{h_{\min} + \tau_{\min}v(t)} \right\rceil-1\right\}$\;
$n_{\mathrm{h, min}} = \max\left\{\hat{n}_{\mathrm{h}}^{\mathrm{prev}} - 1, ~ 0\right\}$\;
\For{$n_{\mathrm{h}} = n_{\mathrm{h, min}}$  \emph{\KwTo} ${n_{\mathrm{h, max}}}$}
{${\hat{s}^{(n_{\mathrm{h}})}_{n_{\mathrm{h}}+2}(t-T_{\mathrm{h}}:t|t) = s_L(t-T_{\mathrm{h}}: t)}$ and ${\hat{v}^{(n_{\mathrm{h}})}_{n_{\mathrm{h}}+2}(t-T_{\mathrm{h}}:t|t) = v_L(t-T_{\mathrm{h}}: t)}$\;
\If{$n_{\mathrm{h}} \neq 0$}
{\For{$i = 0$  \emph{\KwTo} $n_h-1$}{Simulate ${\hat{s}^{(n_{\mathrm{h}})}_{n_{\mathrm{h}}-i+1}(t-T_{\mathrm{h}}:t|t)}$ and ${\hat{v}^{(n_{\mathrm{h}})}_{n_{\mathrm{h}}-i+1}(t-T_{\mathrm{h}}:t|t)}$ with IDM model~\eqref{eqn:IDM_law}\eqref{eqn:IDM_hidm} following vehicle ${n_{\mathrm{h}}-i+2}$ using uniform flow initialization\;}}
Simulate ${\hat{s}^{(n_{\mathrm{h}})}_1(t-T_{\mathrm{h}}:t|t)}$ with IDM model~\eqref{eqn:IDM_law}, \eqref{eqn:IDM_hidm} following vehicle $2$ with real data initialization\;
$J(n_{\mathrm{h}}) = c(n_{\mathrm{h}})\sum_{k = 0}^{T_{\mathrm{s}}}\left(s_1(t-k) - \hat{s}_1^{(n_{\mathrm{h}})}(t-k|t)\right)^2$, where
$$c(n_{\mathrm{h}}) = \left\{\begin{matrix}
1, & {\rm if} \,\,\, n_{\mathrm{h}} = \hat{n}_{\mathrm{h}}^{\mathrm{prev}}~, 
\\
1.5, & {\rm if} \,\,\, n_{\mathrm{h}} \neq \hat{n}_{\mathrm{h}}^{\mathrm{prev}}~,
\end{matrix}\right.$$
}
$\hat{n}_{\mathrm{h}}$ is the index of the minimum of $J$.
\caption{Estimation of Number of Hidden Vehicles}
\label{alg:nh_est}
\end{algorithm}

There are a few details worth mentioning. First, we initialize the state for simulation such that at time ${t-T_{\mathrm{h}}}$, the hidden vehicles are equally spaced with distances
\begin{equation}\label{eqn:sinit_est}
\hat{s}^{(n_\mathrm{h})}_{i+1}(t-T_{\mathrm{h}}) - \hat{s}^{(n_\mathrm{h})}_{i}(t-T_{\mathrm{h}}) = \frac{s_L(t - T_{\mathrm{h}}) - s_1(t - T_{\mathrm{h}})}{n_{\mathrm{h}}+1}~, \qquad i=2, \ldots, n_{\mathrm{h}}+1~,
\end{equation}
and equal velocity
\begin{equation}
\hat{v}_{i}(t-T_{\mathrm{h}}) = \frac{v_1(t-T_{\mathrm{h}}) + v_L(t-T_{\mathrm{h}})}{2}~, \qquad i=2,\ldots, n_{\mathrm{h}}+1~.
\end{equation}
On the other hand, vehicle $1$ is initialized with stored observation data ${\hat{s}_1^{(n_{\mathrm{h}})}(t-T_{\mathrm{h}}) = s_1(t-T_{\mathrm{h}})}$ and\\ ${\hat{v}_1^{(n_{\mathrm{h}})}(t-T_{\mathrm{h}}) = v_1(t-T_{\mathrm{h}})}$, just as vehicle $L=n_{\mathrm{h}}+2$, $\hat{s}^{(n_{\mathrm{h}})}_{n_{\mathrm{h}}+2}(t - T_{\mathrm{h}} : t|t) = s_L(t - T_{\mathrm{h}} : t)$ and $\hat{v}^{(n_{\mathrm{h}})}_{n_{\mathrm{h}}+2}(t - T_{\mathrm{h}} : t|t) = v_L(t - T_{\mathrm{h}} : t)$ are for the V2V data $s_L$, $v_L$.

Second, at the beginning of simulation, there is not enough observation data. Thus, we assume that the vehicles are equally spaced around desired time headway, $\hat{h} = H(v_1)$, and thus, the initial estimation of the number of hidden vehicles is
\begin{equation}
\hat{n}_{\mathrm{h}}^{\mathrm{init}} = \left \lceil \frac{s_{L}(t) - s_{1}(t)}{H\big(v_1(t)\big)} \right \rceil - 1~.
\end{equation}
The estimation is adjusted using Algorithm~\ref{alg:nh_est} as more data is collected. At the beginning, the amount of data is limited, so we use all available data for estimation. After ${t \geq T_{\mathrm{h}}}$, we have abundant data from leading vehicle, so we only use the nearest $T_{\mathrm{h}}$ data. In this paper, we choose ${T_{\mathrm{h}}=\min\{t,\ 23\}}$. Moreover, when comparing the trajectory $\hat{s}^{(n_{\mathrm{h}})}_{1}$ generated from different $n_{\mathrm{h}}$, the cost function $J(n_{\mathrm{h}})$ only compares it to the recorded trajectory $s_1$ within the nearest $T_{\mathrm{s}}$. We choose ${T_{\mathrm{s}}=\min\{t,\ 5\}}$.

Third, we discourage frequent jumps in the value of $n_{\mathrm{h}}$. On one hand, we assume that the number of hidden vehicles should not change over time by more than 1, that is only one vehicle can merge in or out of the ego lane at the same time. On the other hand, we put more penalty on values that are different from the previous estimate $\hat{n}_{\mathrm{h}}^{\mathrm{prev}}$ by $c(n_{\mathrm{h}})$, so $\hat{n}^{\mathrm{prev}}$ is more likely to have smaller cost function, and be chosen in the current step.

\section{Numerical Results}\label{sec:num_result}

In this section, we conduct numerical simulations to compare the four kinds of control algorithms. We demonstrate the benefit of connectivity, prediction, and the generalizability of our algorithms.

\subsection{Simulation Setup}\label{sec:simulation_setup}

First, we introduce the basic setup of our simulation. As is shown in Fig.~\ref{fig:Concepts}, the ego vehicle follows a chain of human driven vehicles. In this paper, we assign real human driving data to the speed trajectories of preceding vehicles. The real data is collected in an experiment where all vehicles were connected. The details of the experiment are described in~\cite{howellexp}.

\begin{figure}[ht]
\centering
\includegraphics[width=0.7\textwidth]{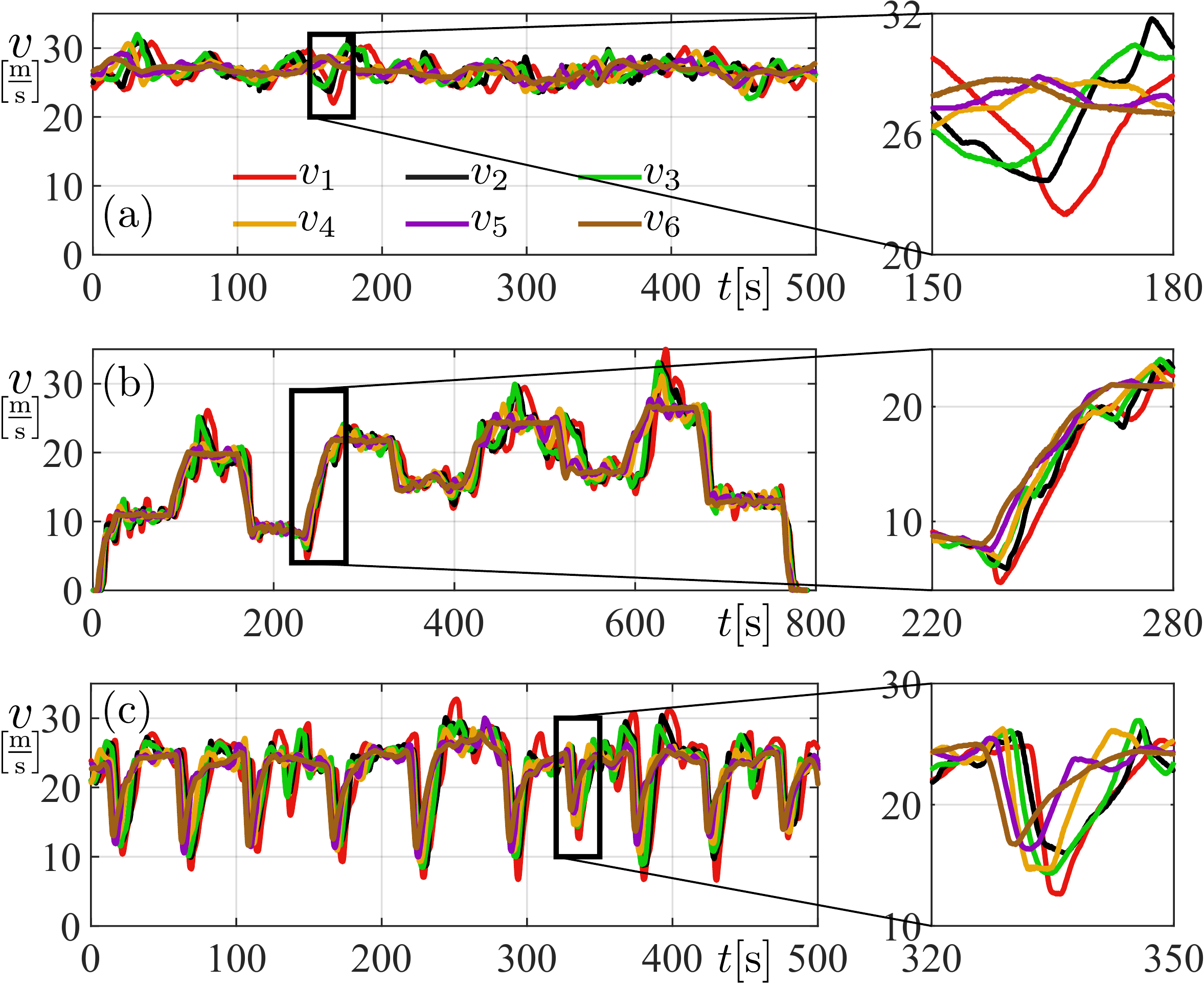}
\caption{Three qualitatively different speed trajectories from experimental data. (a) Free flow profile. (b) Step profile. (c) Congested profile.}
\label{fig:howell}
\end{figure}

In this paper, we consider three kinds of qualitatively different datasets: free-flow, step, and congested, as shown in Fig.~\ref{fig:howell}. In the free-flow profile, the drivers are driving close to the speed limit with little variations. In the step profile, the preceding vehicles accelerate from halt. After reaching a steady-state speed, the vehicles maintain the speed for a while, and then transition to another steady-state speed. In the congested trajectory, the leading vehicle brakes frequently, resulting in consecutive braking by the following vehicles.

In RACC and PACC, the ego vehicle only responds to vehicle $1$. In RCCC and PCCC, connectivity allows the ego vehicle to respond also to the lead vehicle that is chosen to be ranging from $L=2$ to $6$. The nonlinear physical term defined in~\eqref{eqn:f_satu_def} is set to
\begin{equation}
f(v) = 0.0147 + 2.75\times 10^{-4} v^2~.
\end{equation}
The acceleration limits~\eqref{eqn:sat_def}, \eqref{eqn:umax_def}, parameters of range policy and speed policy are shown in Table~\ref{tab:acclim}, \ref{tab:range_policy}, and \ref{tab:rcparam}. 

In RCCC, we fix $\alpha=0.4~\mathrm{[1/s]}$, and the control parameters $\beta_1$, $\beta_L$, $\sigma_L$ are optimized using the method introduced in~\cite{minghao2022stochasticCCC} in each of the three dataset types. The optimal parameters are shown in Table~\ref{tab:rcparam}. The IDM parameters for PCCC are listed in Table.~\ref{tab:idm_param}, and the corresponding MPC parameters are listed in Table.~\ref{tab:mpc_param}. The identification of IDM parameters are described in detail in \ref{sec:idm_identify}.

\begin{table}[ht]
\centering
\begin{minipage}{0.58\linewidth}
\centering
\begin{tabular}{ccccc}
$u_{\min}~\mathrm{[m/s^2]}$ & $m_1~\mathrm{[1/s]}$ & $b_1~\mathrm{[m/s^2]}$ &  $m_2~\mathrm{[1/s]}$  & $b_2~\mathrm{[m/s^2]}$ 
\\ \hline
$-$6   & 0.285 & 2  & $-$0.121 & 4.83
\end{tabular}%
\caption{Acceleration limit of CAV.}
\label{tab:acclim}
\end{minipage}
\begin{minipage}{0.4\linewidth}
\centering
\begin{tabular}{ccccc}
$\tau~\mathrm{[s]}$ & $d~\mathrm{[m]}$ &   $v_{\max}~\mathrm{[m/s]}$ 
\\ \hline
1.67   & 5   & 35 
\end{tabular}%
\caption{Parameters of range policy and speed policy.}
\label{tab:range_policy}
\end{minipage}
\end{table}

\begin{table}[h!]
\centering
\begin{tabular}{cccccc}
                            & Controller Type & $\alpha~\mathrm{[1/s]}$                & $\beta_1~\mathrm{[1/s]}$  & $\beta_L~\mathrm{[1/s]}$  & $\sigma_L~\mathrm{[1/s]}$ \\ \hline
\multirow{2}{*}{Free Flow}  & RACC            & \multirow{2}{*}{0.4} & 0.6617 &        &        \\
                            & RCCC            &                      & 0.3041 & 1.0277 & 5.3372 \\ \hline
\multirow{2}{*}{Step}       & RACC            & \multirow{2}{*}{0.4} & 0.4728 &        &        \\
                            & RCCC            &                      & 0.2163 & 1.1459 & 1.7432 \\ \hline
\multirow{2}{*}{Congested} & RACC            & \multirow{2}{*}{0.4} & 0.4857 &        &        \\
                            & RCCC            &                      & 0.2410 & 0.9895 & 2.4331
\end{tabular}
\caption{Parameters of reactive controllers for $L=6$, i.e.,  ${n_{\mathrm{h}}=4}$.}
\label{tab:rcparam}
\end{table}

\begin{table}[h!]
\centering
\begin{tabular}{ccccccc}
           & $a_0~\mathrm{[m/s^2]}$ & $b_0~\mathrm{[m/s^2]}$ & $\delta~\mathrm{[m]}$ & $\tau~\mathrm{[s]}$ & $d~\mathrm{[m]}$ & $v_{\max}~\mathrm{[m/s]}$ 
\\ \hline
Free flow  & $0.684$  & $2.9693$  & $3.3066$ & $0.7154$ &  $5.0001$  & $36$     
\\
Step       & $2.2868$  & $8.5$  & $3$ & $0.9282$ & $5$ & $32.8682$  
\\
Congested & $2.5732$  & $8.5$  & $4.3393$  & $0.6409$ & $5.067$ & $36$    
\end{tabular}
\caption{IDM parameters identified from free-flow, step and congested datasets.}
\label{tab:idm_param}
\end{table}

\begin{table}[h!]
\centering
\begin{tabular}{cccccccc}
$\Delta t~\mathrm{[s]}$ & $q_{\mathrm{g}}$ & $q_{\mathrm{a}}$  & $q_\epsilon$ & $\tau~\mathrm{[s]}$  & $d~\mathrm{[m]}$ & $\tau_{\min}~\mathrm{[s]}$ & $d_{\min}~\mathrm{[m]}$    
\\ \hline
0.1     & 1  & 960 & $10^6$ & 1.67 & 5 & 0.67   & 3    
\end{tabular}
\caption{Parameters of predictive controllers.}
\label{tab:mpc_param}
\end{table}

\subsection{Benefits of Connectivity}\label{sec:benefit_connectivity}

In this section, we show that connectivity brings great energy benefit. We compare the energy consumption with and without connectivity, and provide an explanation to the observed energy savings by comparing simulated trajectories as well as prediction results.
The optimization problem~\eqref{eqn:mpc_discrete} is formulated in MATLAB with Yalmip~\cite{yalmip} and the quadratic programming problems are solved with Gurobi solver~\cite{gurobi}.

\begin{figure}
\centering
\begin{minipage}{0.45\textwidth}
\centering
    \includegraphics[width=\textwidth]{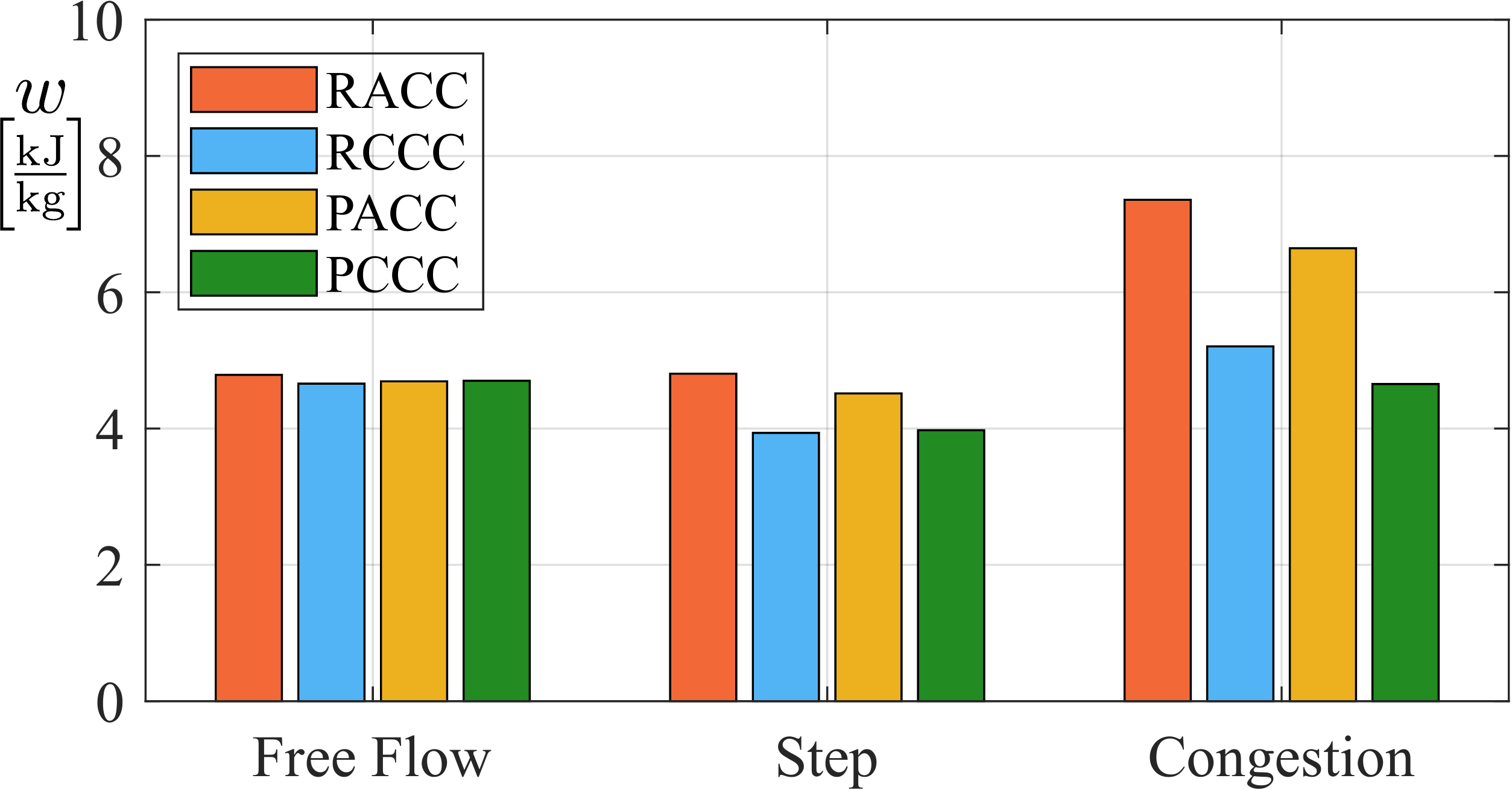}
    \caption{Comparison of energy consumption of RACC, RCCC, PACC, PCCC.}
    \label{fig:cc_comp_all_nh4}
\end{minipage}
\hspace{0.05\textwidth}
\begin{minipage}{0.45\textwidth}
\centering
\includegraphics[width=\textwidth]{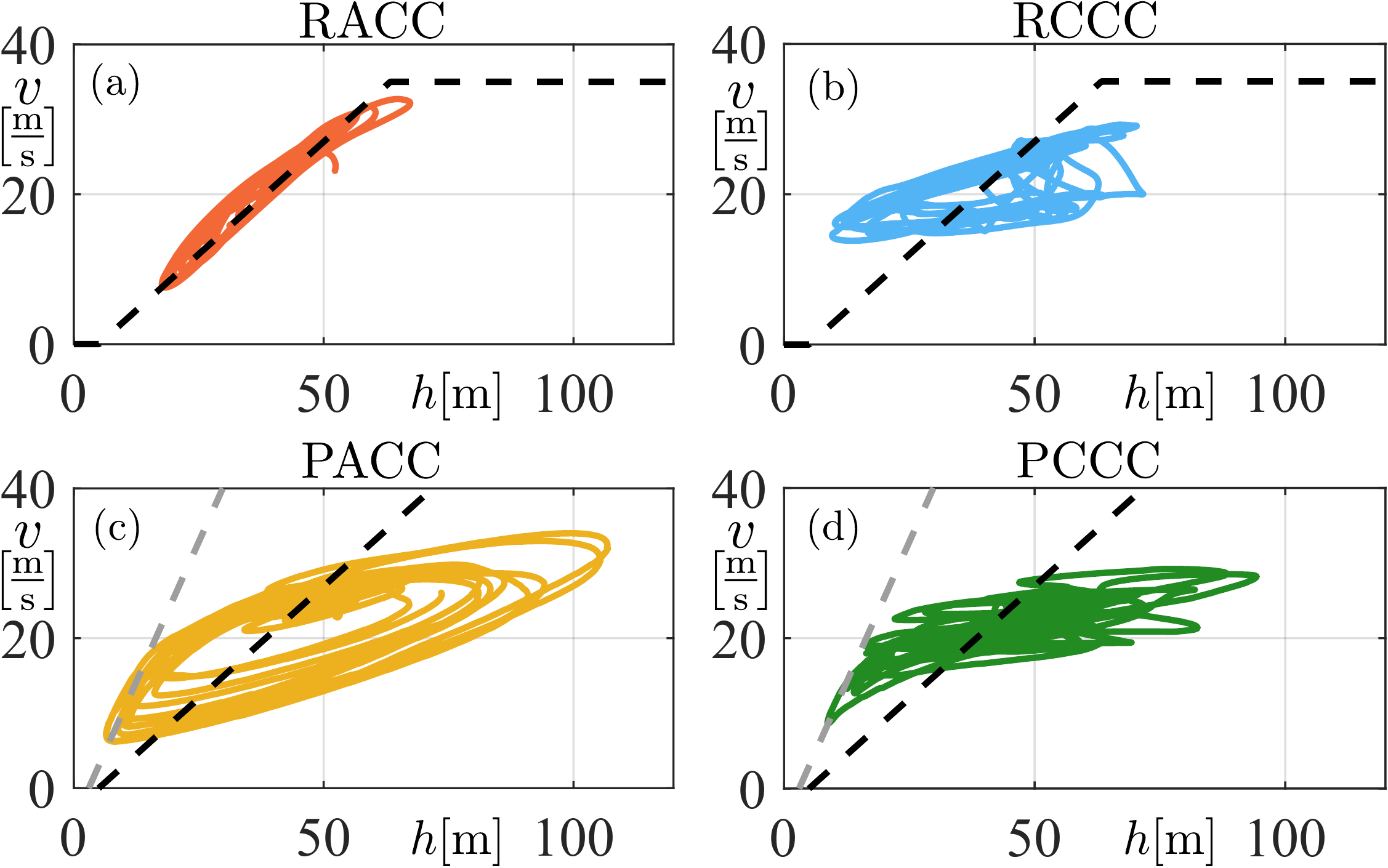}
\caption{Phase portrait for RACC, RCCC, PACC, PCCC. Desired range policy is plotted in black dashed line and the safety constraints for predictive controllers are indicated by grey dashed line.}
\label{fig:hvplot}
\end{minipage}
\end{figure}

In three traffic scenarios, we compare the energy consumption of the four controllers we introduced above: RACC, RCCC, PACC, and PCCC, as shown in Fig.~\ref{fig:cc_comp_all_nh4}. 
In free flow scenario, the speed variations of preceding vehicles are small, so there is little difference on the energy efficiency of the four controllers. While in the step and congested scenarios, controllers with connectivity~(RCCC and PCCC) save significant amount of energy compared to those without connectivity~(RACC and PACC). Compared to RACC, RCCC saves $18.1\%$ energy in step scenario and $29.2\%$ energy in congested scenario. Compared to PACC, PCCC saves $12.0\%$ energy in step scenario and $30.0\%$ energy in congested scenario. It is also worth noting that RCCC and PCCC achieve similar energy consumption in free-flow and step scenarios, but PCCC consumes $11.9\%$ less energy compared to RCCC in congested scenario.

\begin{figure}[h!]
\centering
\begin{minipage}{.45\textwidth}
  \centering
  \includegraphics[width=\linewidth]{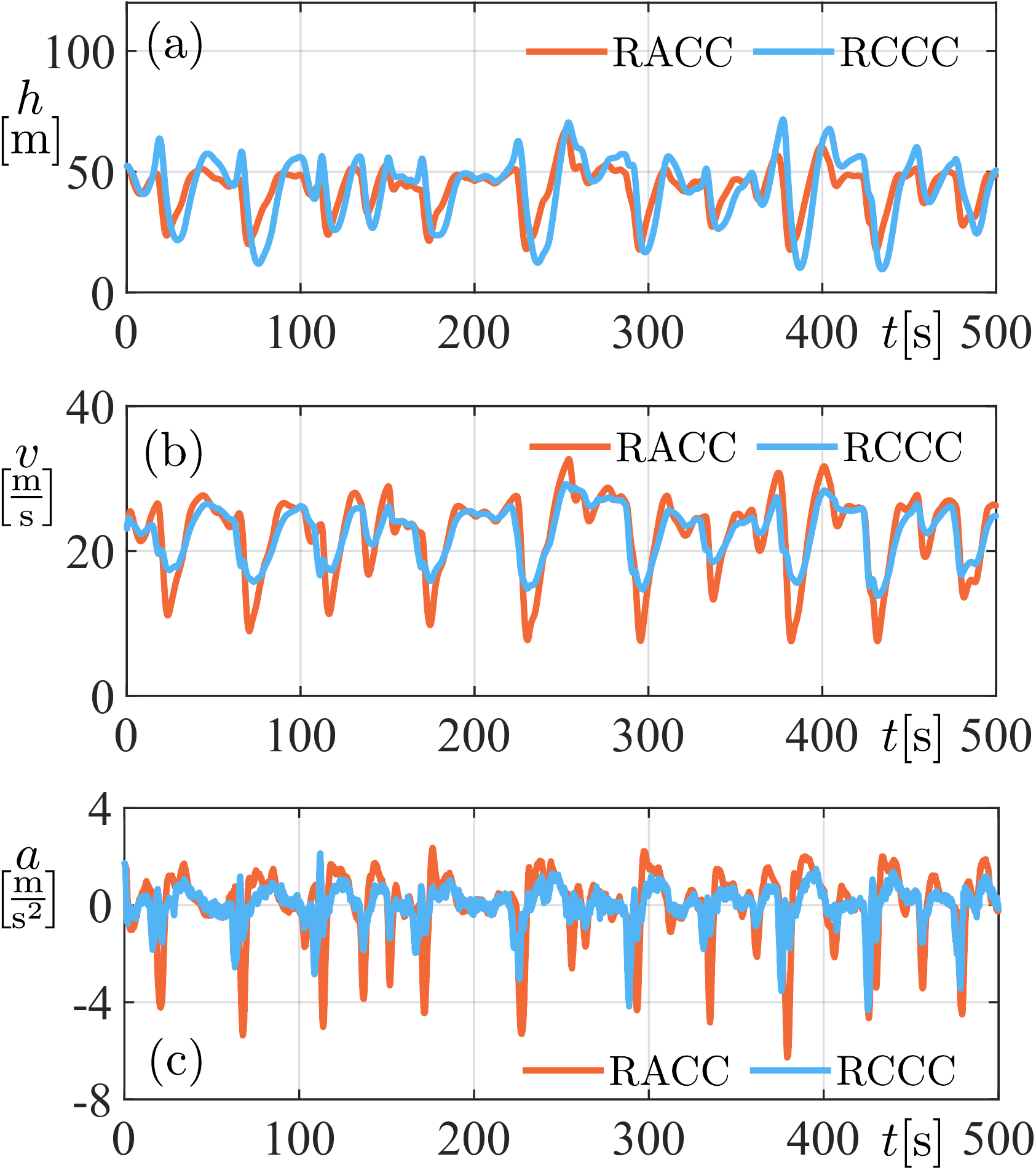}
  \caption{Comparison of RACC and RCCC in congested scenario for ${L=6}$, i.e., ${n_{\mathrm{h}}=4}$. (a) Headway; (b) Speed; (c) Acceleration.}
  \label{fig:racc_rccc_traj}
\end{minipage}%
\hspace{0.05\linewidth}
\begin{minipage}{.45\textwidth}
  \centering
  \includegraphics[width=\linewidth]{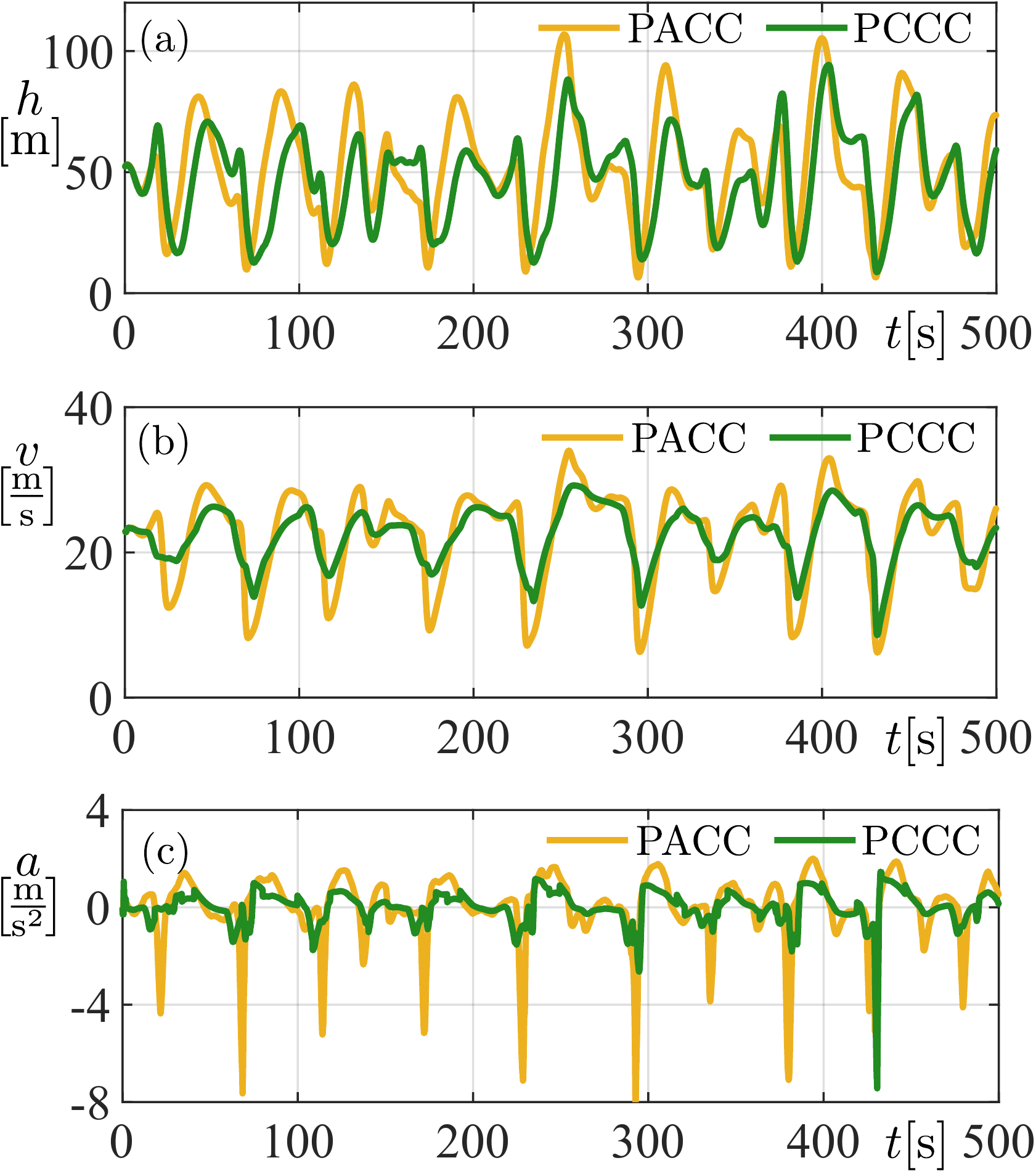}
  \caption{Comparison of PACC and PCCC in congested scenario for ${L=6}$, i.e.,  ${n_{\mathrm{h}}=4}$. (a) Headway; (b) Speed; (c) Acceleration.}
  \label{fig:pacc_pccc_traj}
\end{minipage}
\end{figure}

In Fig.~\ref{fig:hvplot}, we plot the trajectories in the $(h,v)$-plane. This demonstrates that reactive controllers, especially RACC, adhere to the pre-defined range policy, while predictive controllers have more degrees of freedom in deviation from the nominal headway-speed relationship. In addition, we plot the trajectories of headway, velocity and acceleration in congested scenario in Fig.~\ref{fig:racc_rccc_traj}, \ref{fig:pacc_pccc_traj}. Without connectivity, vehicles suffer from abrupt brakings and accelerations, which leads to excessive energy consumption. With information from V2V connectivity, both reactive and predictive controllers can obtain smaller speed variations while maintaining reasonable headway. We also observe that predictive controllers tend to keep larger headways compared to reactive controllers.

\begin{figure}[h!]
\centering
\includegraphics[width=0.85\textwidth]{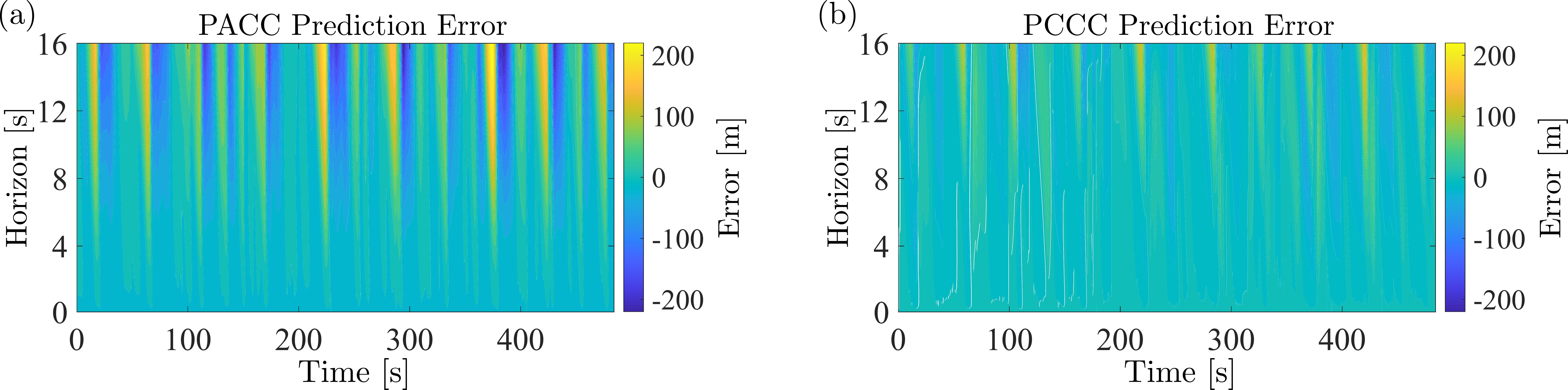}
\caption{(a)~Prediction error for vehicle $1$ of PACC based on constant speed assumption. (b)~Prediction error of PCCC based on car-following model.}
\label{fig:pred_err_pacc_pccc}
\end{figure}

For predictive controllers, we can directly show the improvement of prediction due to additional information from V2V connectivity. Figure~\ref{fig:pred_err_pacc_pccc} shows the error between the prediction and the ground truth 
\begin{equation}
\mathrm{error}(t, k) = \hat{s}_1(k|t) - s_1(t+k)~.
\end{equation}
for PACC and PCCC. In both cases the accuracy is similar up to about 4 seconds time horizon. However, with additional information from V2V connectivity, PCCC can achieve better prediction on a larger horizon.

While constant speed assumption is used in both PACC and PCCC predictions,  PACC assumes constant speed for vehicle $1$ while PCCC assumes constant speed for vehicle $L$ in the distance and predicts the motion of vehicle 1 by a car-following model. The motion of vehicle 1 in the near future is affected by the {\em past} motion of vehicle $L$, hence the improvement of prediction when using data from vehicle $L$ by connectivity.

\begin{figure}[h!]
    \centering
    \includegraphics[width=0.81\textwidth]{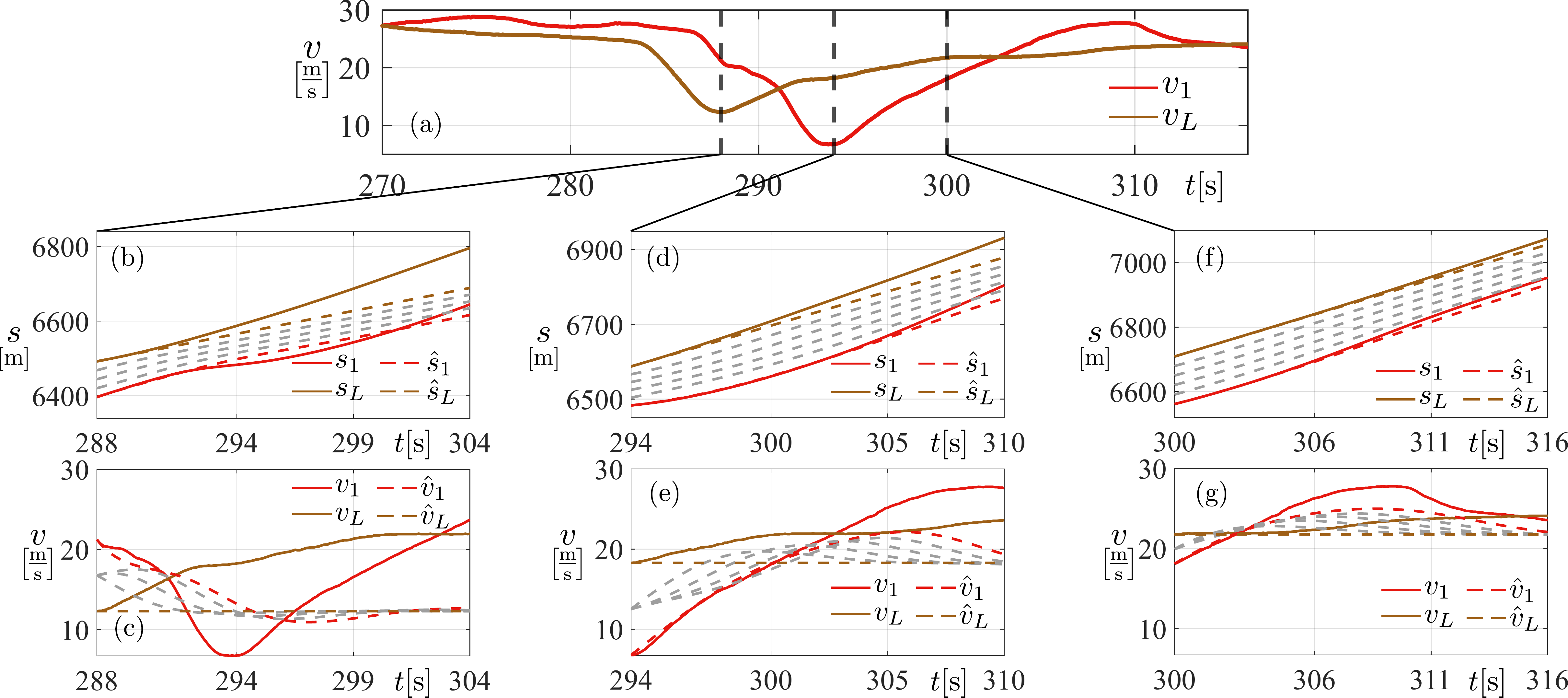}
    \caption{Comparison of IDM prediction and the ground truth. Dashed curves show the predicted future trajectories, while solid curves show the trajectory data collected from experiments.}
    \label{fig:pccc_predshow}
\end{figure}

Figure~\ref{fig:pccc_predshow} further illustrates the motion prediction in a deceleration-acceleration cycle. We show the motion predictions at ${t=288~[\mathrm{s}]}$ (during deceleration), ${t=294~[\mathrm{s}]}$ (at the end of deceleration), ${t=300~[\mathrm{s}]}$ (during acceleration). At each time instance, we plot the ground truth positions and speeds of vehicles $1$ and $L$ by solid lines, and the corresponding predictions by dashed lines. Meanwhile we show the prediction of hidden vehicles by grey dashed lines. At ${t=288~[\mathrm{s}]}$ the IDM model is able to predict the deceleration at the initial ${4~[\mathrm{s}]}$ horizon, but fails to predict the harsh deceleration and the acceleration afterwards; see panels (b), (c). However, when the leading vehicle finishes braking and starts mildly accelerating, the IDM model is able to capture the car-following behavior and achieve a good prediction of speed for more than 6 seconds, see panels (d), (e) for ${t=294~[\mathrm{s}]}$ and panels (f), (g) for ${t=300~[\mathrm{s}]}$.

\begin{figure}[h!]
    \centering
    \includegraphics[width=0.98\textwidth]{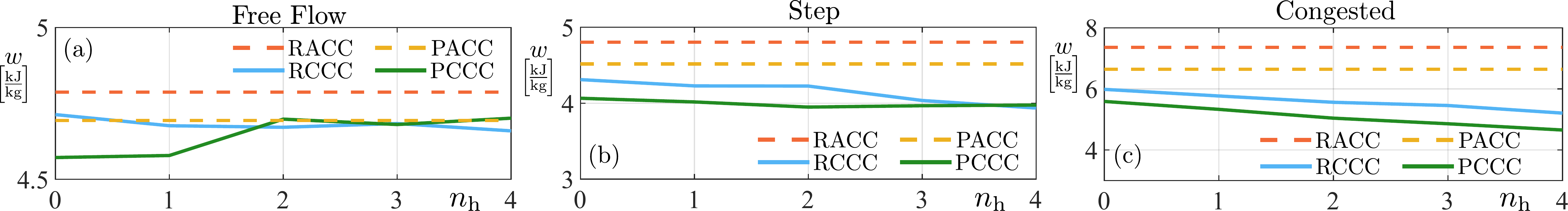}
    \caption{Comparison of the energy consumption of RACC, PACC, RCCC and PCCC as a function of the number of hidden vehicles for the three traffic datasets in Fig~\ref{fig:howell}.}
    \label{fig:rpccc_comp_nhall}
\end{figure}

In practice, the number of hidden vehicles is unknown.  Fig.~\ref{fig:rpccc_comp_nhall} shows the sensitivity to the number of hidden vehicles~$n_{\mathrm{h}}$ in free-flow, step, and congested scenarios.
In general, with connectivity, significant energy can be saved in all three scenarios. However, in free flow scenarios, connection to vehicles farther ahead associated with larger $n_{\mathrm{h}}$ may increase the energy consumption for PCCC. While in other scenarios, connecting to farther vehicles helps reduce the energy consumption. Especially in congested scenario, connecting to a vehicle in the distance~(${n_{\mathrm{h}}=4}$) saves $13.0\%$ energy compared to connecting to a nearby vehicle~(${n_{\mathrm{h}}=1}$) in case of RCCC, and $16.8\%$ in case of PCCC.

However, with more hidden vehicles involved in the prediction, more uncertainties are introduced. For example, in the free-flow scenario, see Fig.~\ref{fig:howell}(a), vehicles $4$, $5$ and $6$ are running with small speed fluctuation, but the speed fluctuation is significantly larger for vehicle 3. If the source of fluctuation is not observable, i.e., vehicle 2 and 3 are hidden vehicles, it is hard to make a precise predictions using car-following model. In conclusion, when the speeds of the leading vehicle and the preceding vehicle are highly correlated, connecting to vehicles farther ahead usually helps saving energy until the increasing uncertainty undermines the correlation and causes more energy consumption.

\section{Conclusions and Future Work}\label{sec:concl}

In this paper, we proposed a framework for longitudinal control design for automated vehicles driving in mixed traffic consisting of connected and non-connected vehicles. The longitudinal controllers included reactive controllers, where an explicit feedback law was assigned, and predictive controllers, where the control input was optimized in receding horizon fashion, according to the predicted future motion of preceding vehicles.

The controllers realized adaptive cruise control and connected cruise control. In the latter case beyond-line-of-sight information was obtained using vehicle-to-vehicle~(V2V) communication. With lean penetration of connected vehicles in the traffic, various techniques were applied to improve the energy efficiency. In reactive controllers, the controller parameters were optimized according to the observed data. In predictive controllers, the number of hidden vehicles was estimated online, and a car-following model was applied to predict the motion of preceding vehicles.

We conducted extensive simulations based on real human driver data for various driving scenarios. We showed that even lean penetration of connectivity can bring significant energy benefits, with both reactive and predictive controllers, in all driving scenarios. The influence of the number of hidden vehicles was also studied: connection to vehicles farther in the distance usually brings additional energy benefits until the increasing uncertainty undermines these benefits.

The proposed control framework can accommodate various engineering specifications, for example, different implementations of reactive control law, estimation of the number of hidden vehicles, or motion predictors. The framework is also applicable to various kinds of vehicles including internal-combustion engine vehicles, electric vehicles, or hybrid electric vehicles. While, the design framework in this paper only considers longitudinal motion in single lane, it is promising to consider both longitudinal and lateral motion on multiple lanes in future research.

\appendix
\section{Safety Margin}\label{sec:margin}

In \eqref{eqn:safety_ineq} we introduced the safety margin $d_{\mathrm{margin}}$ to compensate for the uncertainty in the motion of vehicle $1$ immediately in front of the automated vehicle~\cite{ard2020microsimulation}. Considering the randomness of vehicle motion, we use the mean of the random process as predictions about the motion of preceding vehicles. Let $s_1(k)$, $v_1(k)$, $a_1(k)$ denote the position, velocity and acceleration of vehicle 1 at time moments ${k=0,\ldots, T-1}$. Using the notation ${x_1(k) = \big[s_1(k), \ v_1(k)\big]^\top}$, \eqref{eqn:disc_cardyn} can be written as
\begin{equation}
x_1(k+1) = A\, x_1(k) + B\, a_1(k)~,
\end{equation}
where
\begin{equation}
A = 
\begin{bmatrix}
1 & \Delta t 
\\ 
0 & 1
\end{bmatrix}~, 
\qquad B = 
\begin{bmatrix}
\frac{1}{2}\Delta t^2 
\\ 
\Delta t
\end{bmatrix}~.
\end{equation}
Then ${x_1 := \big[x_1^\top(1), \ x_1^\top(2), \ \cdots, \ x_1^\top(T-1)\big]^\top}$ is given by
\begin{equation}
x_1 = \hat{A}\, x_1(0) + \hat{B}\, a_1~,
\end{equation}
where ${a_1 = \big[a_1(0), a_1(1), \ \cdots, \ a_1(T-2)\big]^\top}$, and
\begin{equation}
\hat{A} = 
\begin{bmatrix}
A 
\\ 
A^2 
\\ 
\vdots 
\\ 
A^{T-1}
\end{bmatrix}~, 
\qquad 
\hat{B} = 
\begin{bmatrix}
B & 0 & \cdots & 0 
\\ 
A B & B & \cdots & 0 
\\ 
\vdots & \vdots & \ddots & \vdots 
\\ 
A^{T-2}B & A^{T-3}B & \cdots & B
\end{bmatrix}~.
\end{equation}

We split deterministic and stochastic part of the acceleration profile as 
\begin{equation}
a_1 = \hat{a}_1 + \tilde{a}_1~, 
\end{equation}
where the profile $\hat{a}_1$ is provided by the deterministic car-following model and, for simplicity, we consider the uniform noise profile
\begin{equation}
\tilde{a}_1 = 
\big[1,\ 1,\ \cdots, \ 1 \big]^\top e_{a_1}~, 
\qquad 
e_{a_1} \sim \mathcal{N}\Big(0, \ \sigma_{a_1}^2\Big)~.
\end{equation}
These result in
\begin{equation}
\hat{B}\, a_1 = \hat{B}\,\overline{a}_1 +  \hat{B}\,\tilde{a}_1 = \hat{B}\,\overline{a}_1 + 
\underbrace{
\begin{bmatrix} 
B 
\\
A B + B 
\\
\vdots 
\\
A^{T-2} B + A^{T-3} B + \cdots B 
\end{bmatrix}
}_{\hat{\hat{B}}}
e_{a_1}~. 
\end{equation}  
Thus, we can derive the distribution of $x_1$ as
\begin{equation}
x_1 \sim \mathcal{N} \Big( \hat{A}\,x_1(0) + \hat{B}\, \overline{a}_1, \ \hat{\hat{B}} \hat{\hat{B}}^\top \sigma_{a_1}^2 \Big)~.
\end{equation}
Similar to the acceleration we may also split the state as
\begin{equation}
x_1 = \hat{x}_1 + \tilde{x}_1~, 
\end{equation}
where
\begin{equation}
\hat{x}_1 = \hat{A}\,x_1(0) + \hat{B}\, \overline{a}_1~,    
\end{equation}
represents the deterministic part as a result of the deterministic car-following model, and 
\begin{equation}
\tilde{x}_1 \sim \mathcal{N}\Big(0, \hat{\hat{B}} \hat{\hat{B}}^\top \sigma_{a_1}^2\Big)~,
\end{equation}
represents the stochastic part. Utilizing the notations ${\tilde{x}_1(k) = \big[\tilde{s}_1(k), \ \tilde{v}_1(k) \big]^\top}$ and ${\tilde{x}_1 = \big[ \tilde{x}_1^\top(1), \ \tilde{x}_1^\top(2), \ \cdots, \ \tilde{x}_1^\top(T-1) \big]^\top}$, the calculations above yield
\begin{equation}
\tilde{s}_1(k) \sim \mathcal{N} \Big( 0, \sigma_{s_1}^2(k)\Big)~, \qquad 
\sigma_{s_1}^2(k) = b_k\, \sigma_{a_1}^2~.   
\end{equation}
where $b_k$ is the element of matrix $\hat{\hat{B}}\hat{\hat{B}}^{\top}$ at the ${(2k-1)}$-th row and ${(2k-1)}$-th colomn.

\begin{figure}[h!]
    \centering
    \includegraphics[width=0.5\textwidth]{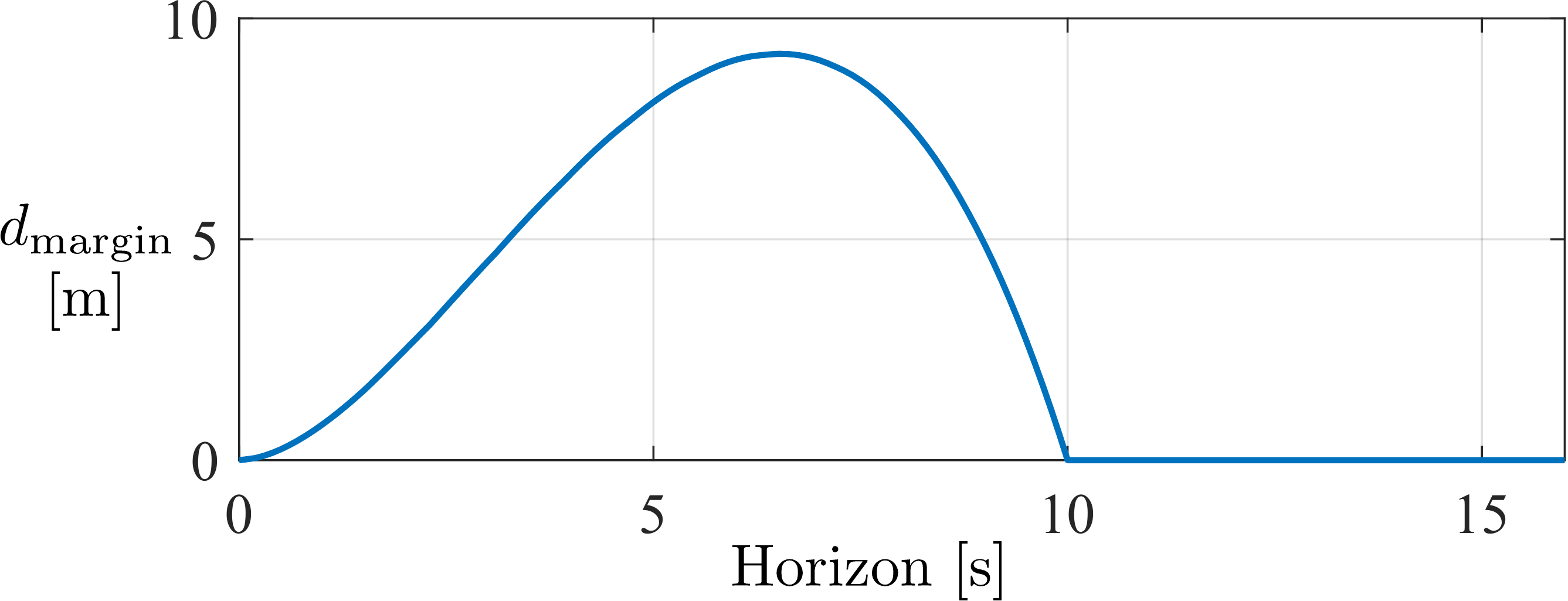}
    \caption{Safety margin within prediction horizon.}
    \label{fig:chance-const}
\end{figure}

In order to compensate for the uncertainty in $s_1$ we enforce the probabilistic safety constraint
\begin{equation}\label{eqn:safe_prob}
\mathbb{P}\Big[s_1(k) - s(k) - l - H_{\min}\big(v(k)\big) \geq 0 \Big] \geq \alpha(k)~,
\end{equation}
which can be rewritten as
\begin{equation}
\mathbb{P}\Big[\tilde{s}_1(k) \geq s(k) + H_{\min}\big(v(k)\big) + l - \hat{s}_1(k) \Big] \geq \alpha(k)~.
\end{equation}
Using the cumulative density function ${\Phi(z) = \frac{1}{\sqrt{2\pi}}\int_{-\infty}^{z}e^{-\frac{t^2}{2}}\mathrm{d}t}$ of the standard Gaussian distribution we obtain
\begin{equation}
 1 - \Phi\left(\frac{s(k) + H_{\min}\big(v(k)\big) + l - \hat{s}_1(k)}{\sigma_{s_1}(k)}\right) \geq \alpha(k)~,
\end{equation}
and exploiting that $1 - \Phi(\alpha) = \Phi(-\alpha)$ this leads to
\begin{equation}
 \hat{s}_1(k) - s(k) - l - H_{\min}\big(v(k)\big) - \sigma_{s_1}(k) \, \Phi^{-1}\big(\alpha(k)\big) \geq 0~.
\end{equation}
Therefore, we define the safety margin
\begin{equation}
d_{\mathrm{margin}}(k) = \max\Big\{0,\, \sigma_{s_1}(k)\, \Phi^{-1}\big(\alpha(k)\big)\Big\}~.
\end{equation}
In this paper, we choose a linearly decreasing $\alpha(k)$ with ${\alpha(1) = 0.99}$ and ${\alpha(K)  =0.5}$, where ${K\Delta t = 10~\mathrm{[s]}}$, to gradually loosen the probabilistic constraint within the prediction horizon. The corresponding safety margin is plotted in Fig.~\ref{fig:chance-const} as a function of time.

\section{IDM parameter identification}\label{sec:idm_identify}

The IDM parameters are obtained via solving the optimization problem
\begin{equation}\label{eqn:idm_param_opt}
\begin{aligned}
\min_{a_0, b_0, \delta, \tau, h_{\mathrm{st}}, v_{\max}}\quad &\frac{1}{N_{\mathrm{v}}}\sum_{i=1}^{N_{\mathrm{v}}}\sqrt{\frac{1}{N_{\mathrm{s}}}\sum_{k=1}^{N_{\mathrm{s}}}\left(\hat{h}_i(k) - h_{i}(k)\right)^2}~, 
\\
\mathrm{s.t.}\quad& 0.1 \leq a_0 \leq  4~, 
\\
& 0.1 \leq b_0 \leq 8.5~, 
\\
& 3 \leq \delta \leq 5~, 
\\
& 0.1 \leq \tau \leq 4~, 
\\
& 5 \leq d \leq 10~, 
\\
& 30 \leq v_{\max} \leq 36~.
\end{aligned}
\end{equation}
In each dataset, there are $6$ human drivers and the cost function averages the performance of IDM model on all ${N_{\mathrm{v}}=5}$ vehicles. For each vehicle, $h_i$ represents the data while $\hat{h}_i$ is obtained from simulations (using the same initial conditions as those in the data) for total time length $N_{\mathrm{s}}$. We set some lower bounds and upper bounds on the IDM parameters considering their physical meanings. We use NOMAD optimizer~\cite{NOMAD} to solve the optimization problem~\eqref{eqn:idm_param_opt}. The resulting IDM parameters are shown in Table~\ref{tab:idm_param}.

\section*{Acknowledgements}
This research was supported by the University of Michigan’s Center for Connected and Automated Transportation through 
the US DOT grant 69A3551747105.

\bibliographystyle{elsarticle-num} 
\bibliography{2021_TRC_MHS}

\end{document}